\documentclass[jkps,preprint,fleqn,showpacs,showkeys]{revtex4}
\usepackage{graphicx}
\usepackage{amssymb}
\usepackage{amsmath}
\usepackage{xspace}
\usepackage{xcolor}

\newcommand{\pt}{\mbox{$p_\mathrm{T}$}\xspace}

\begin{document}
\setcounter{page}{0}
\title[]{Exploring an image-based $b$-jet tagging method \\ 
using convolution neural networks}
\author{HanGil \surname{Jang}}
\affiliation{Department of Physics, Pusan National University, Busan 46241}
\affiliation{Department of Convergence Security, Kangwon National University, Chuncheon, 24341}
\author{SangHoon \surname{Lim}}
\email{shlim@pusan.ac.kr}
\affiliation{Department of Physics, Pusan National University, Busan 46241}
\affiliation{Extreme Physics Institute, Pusan National University, Busan 46241}


\begin{abstract}
Jet flavor tagging, the identification of jets originating from $c$-quarks, $b$-quarks, and other quarks (light quarks and gluons), is a crucial task in high-energy heavy-ion physics, as it enables the investigation of flavor-dependent responses within the hot and dense nuclear medium produced in heavy-ion collisions. 
Recently, several methods based on deep learning techniques, such as deep neural networks and graph neural networks, have been developed. These deep-learning-based methods demonstrate significantly improved performance compared to traditional methods that rely on track impact parameters and secondary vertices.
In the tagging algorithms, various properties of jets and constituent charged particles are used as input parameters. We explore a new method based on images surrounding the primary vertex, utilizing charged particles within the jet cone, which can be measured using a silicon tracking system. 
For this initial experimental study, we assume the ideal performance of the tracking system.
To analyze these images, we employed convolutional neural networks.
The image-based flavor tagging method shows an 80--90\% $b$-jet tagging efficiency for jets in the transverse momentum range from 20 to 100 GeV/$c$. This approach has the potential to significantly improve the accuracy of jet flavor tagging in high-energy nuclear physics experiments.

\end{abstract}



\maketitle

\section{Introduction}

Heavy-ion collisions at high energies provide a unique environment for studying the quark-gluon plasma (QGP), a deconfined state of quarks and gluons believed to have existed for microseconds after the Big Bang \cite{PHENIX:2004vcz, STAR:2005gfr, PHOBOS:2004zne, BRAHMS:2004adc, ALICE:2022wpn}. Among the various probes of the QGP, the study of heavy-flavor jets, particularly those originating from bottom quarks ($b$-jets), offers essential insight into parton energy loss mechanisms and the transport properties of heavy quarks traversing the medium \cite{Gyulassy:1990ye, Dong:2019byy}. Due to their large mass and long lifetime, bottom quarks exhibit distinct energy loss patterns compared to charm and light quarks, making $b$-jet measurements a powerful tool for testing theoretical models on jet-medium interactions.

The identification of $b$-jets exploits specific features such as the presence of secondary vertices and displaced tracks resulting from the relatively long lifetime of B hadrons. Traditional $b$-tagging methods rely on variables like track impact parameters and reconstructed secondary vertices \cite{CMS:2012feb, ATLAS:2015thz}. However, with the increasing complexity of collision environments, especially in heavy-ion collisions, machine learning (ML) and deep learning (DL) techniques have become indispensable for enhancing jet flavor classification performance \cite{Mondal:2024nsa}. Algorithms that utilize convolutional neural networks (CNNs) \cite{oshea2015} and recurrent neural networks (RNNs) \cite{schmidt2019}, considering a complex correlation between input variables, have shown enhanced performance compared to classical methods \cite{Bols:2020bkb, ATLAS:2017gpy}. 

Later, a deep neural network (DNN) was developed that integrates the outputs of several independently optimized low-level algorithms as its input features. These low-level algorithms are designed to extract specific signatures associated with heavy-flavor jets using tracking information as well as variables from machine-learning-based taggers \cite{Guest:2016iqz, ATLAS:2022qxm}. A recent development utilizing a graph neural network (GNN) \cite{battaglia2018} employs a single neural network that directly processes track information, along with select jet-level features, as inputs. This approach eliminates the reliance on separate flavor tagging algorithms, enabling the optimization of all components of the tagging process through a single, unified training \cite{ATLAS:2022rkn}.

These algorithms are based on kinematic variables of jets and their constituent tracks. Building on these advances, we propose an image-based approach to $b$-jet classification using a CNN framework, by transforming track information into spatially structured images centered on the primary vertex. Assuming an ideal tracking detector with perfect spatial resolution and efficiency, this study aims to assess the feasibility and potential of such image-driven techniques for heavy-flavor jet tagging in the context of heavy-ion physics. This paper is organized as follows: Section~\ref{sec:sim} describes the simulation setup and input data used in the study, and Section~\ref{sec:nn} describes the neural network architecture and details of functions. Section~\ref{sec:results} provides the results of the experiments, and a summary follows in Section~\ref{sec:summary}.

\section{Simulation setup}
\label{sec:sim}
We utilize the PYTHIA8 Monte-Carlo event generator~\cite{Bierlich:2022pfr} to simulate proton-proton collisions $\sqrt{s}$ = 5 TeV and obtain jet samples.
We run a few million events for each light-jet, $c$-jet, and $b$-jet samples with an option of ``HardQCD:all'', ``HardQCD:ccbar'', and ``HardQCD:bbbar'', respectively.
To enrich high transverse momentum (\pt) jets, the ``PhaseSpace:pTHatMin = 20.0'' option is included.

For each event, jets are reconstructed using final-state particles based on the anti-$k_T$ algorithm~\cite{Cacciari:2008gp} with a resolution parameter $\Delta R = 0.4$ ($\Delta R \equiv \sqrt{(\Delta \eta)^{2} + (\Delta \varphi)^{2} }$). 
Jets are required to have a \pt greater than 20 GeV$/c$ and an absolute pseudorapidity within 0.5 to focus on the acceptance of the ALICE experiment.
A truth-level jet flavor tagging is performed by all the history of fragmentation and decay in PYTHIA8.
If a $b$-hadron is found within the jet cone ($\Delta R = 0.4$), the jet is tagged as $b$-jet.
If only $c$-hadron (no $b$-hadron) is found within the jet cone, the jet is tagged as $c$-jet, and all other jets are tagged as light-jets.
Jets for each flavor are selected from each simulation set of different ``HardQCD'' options.

Jet images are produced with charged particles within the jet cone, based on the fact that a silicon tracking system measures charged particles.
We consider both ideal and realistic cases of the tracking system regarding tracking efficiency and momentum measurement.
For the ideal case, all charged particles within the jet cone are used with the truth momentum information.
For the realistic case, we refer to the momentum resolution and the tracking efficiency from the expected performance of the ITS3 detector in the ALICE experiment \cite{ALICE:2022wwr, The:2890181}.
The momentum resolution is 3--5\%, and the tracking efficiency is 80\% at $p_{T}=0.1~\mathrm{GeV}/c$ and gradually increases and reaches 95\% at $p_{T}=0.5~\mathrm{GeV}/c$.
Based on the momentum resolution as a function of $\pt$, we apply a Gaussian smearing to the $p_x$ and $p_y$ values, thereby modifying the directional vector of the tracks.
Additionally, the tracking efficiency is applied by generating a random number between 0 and 1 and removing tracks if the random number exceeds the efficiency value.
For each jet, two types of two-dimensional images in the $x-y$ plane are created as follows.
\begin{itemize}
    \item Line image: each constituent charged particle is drawn as a straight line based on its production vertex position and momentum
    \item Dot image: Intersections of at least two lines in the line image are drawn as dots
\end{itemize}

\begin{figure}[htb]
\includegraphics[width=0.49\textwidth]{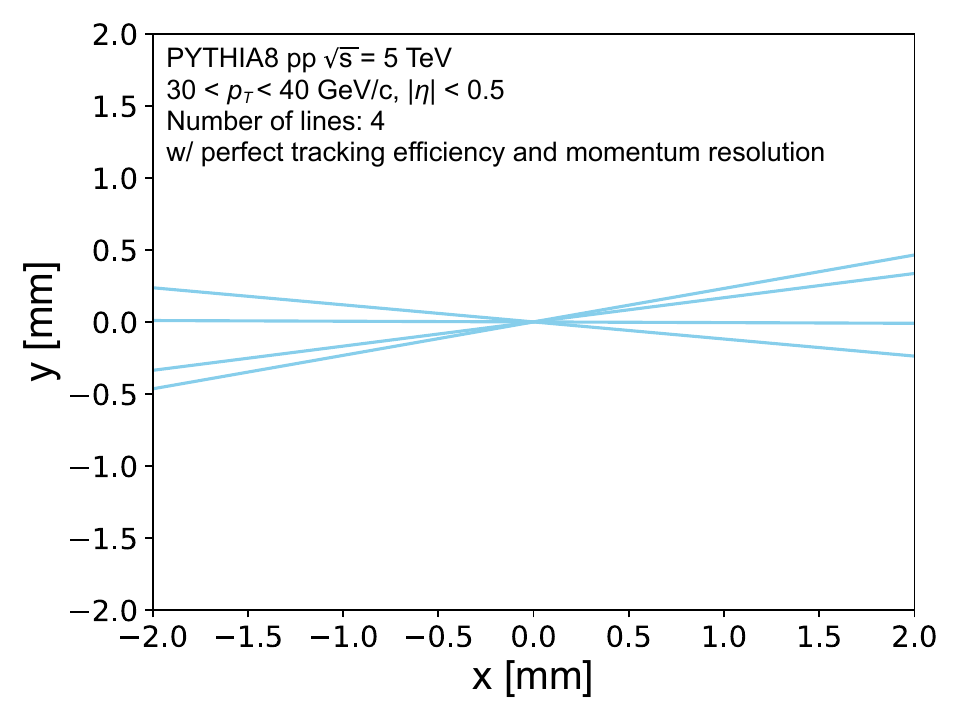}
\includegraphics[width=0.49\textwidth]{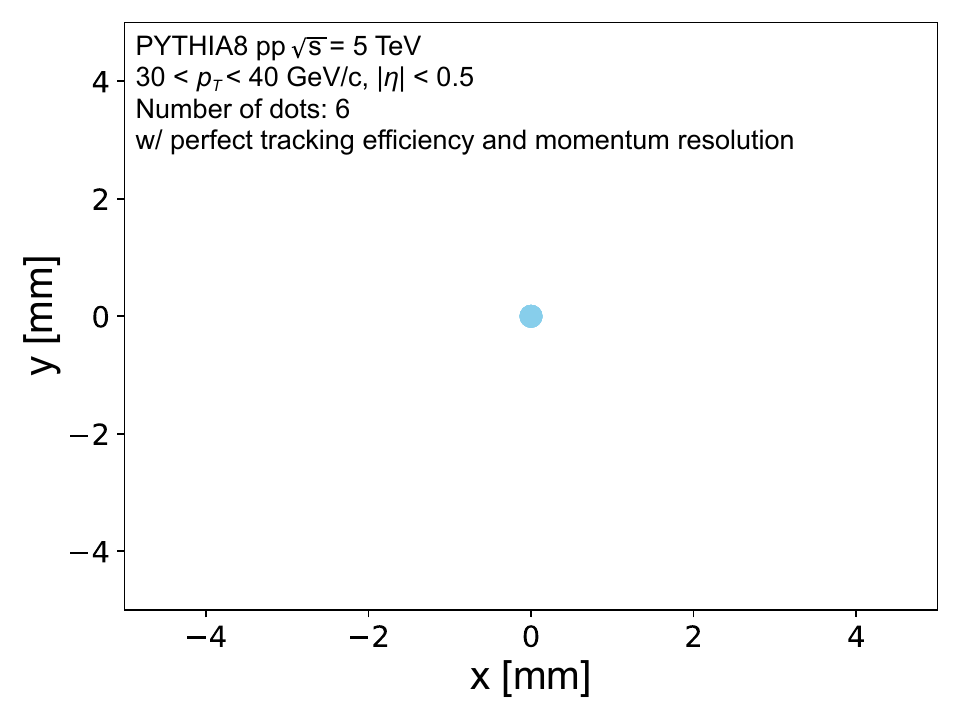}
\caption{Example 2D images of a light-jet with lines (left) and dots (right) with perfect tracking efficiency and momentum resolution.}
\label{fig:image_ljet}
\end{figure}

\begin{figure}[htb]
\includegraphics[width=0.49\textwidth]{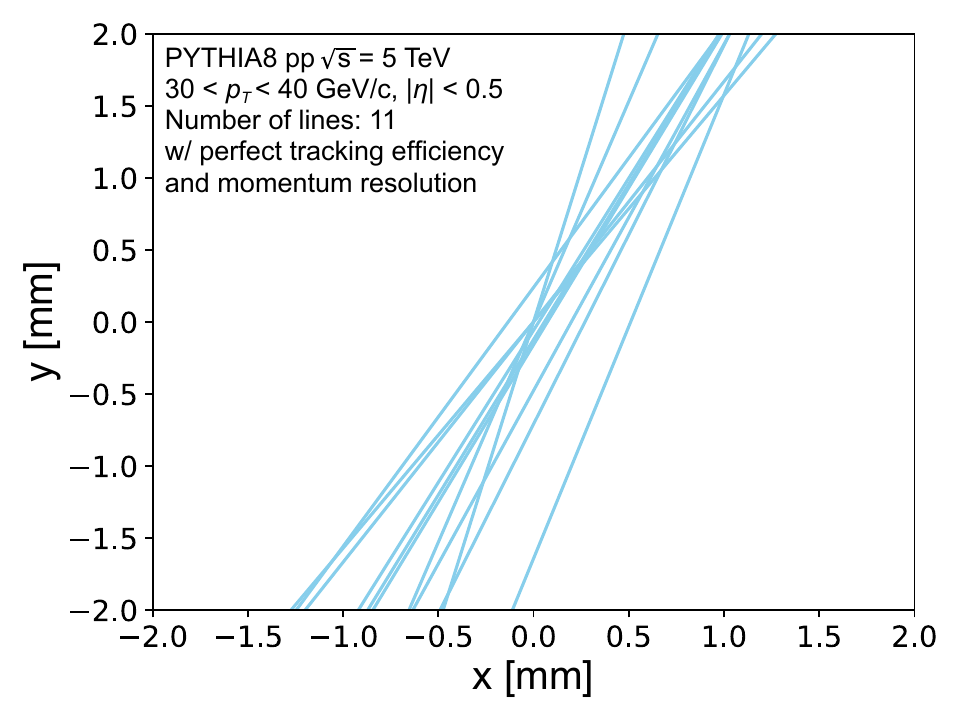}
\includegraphics[width=0.49\textwidth]{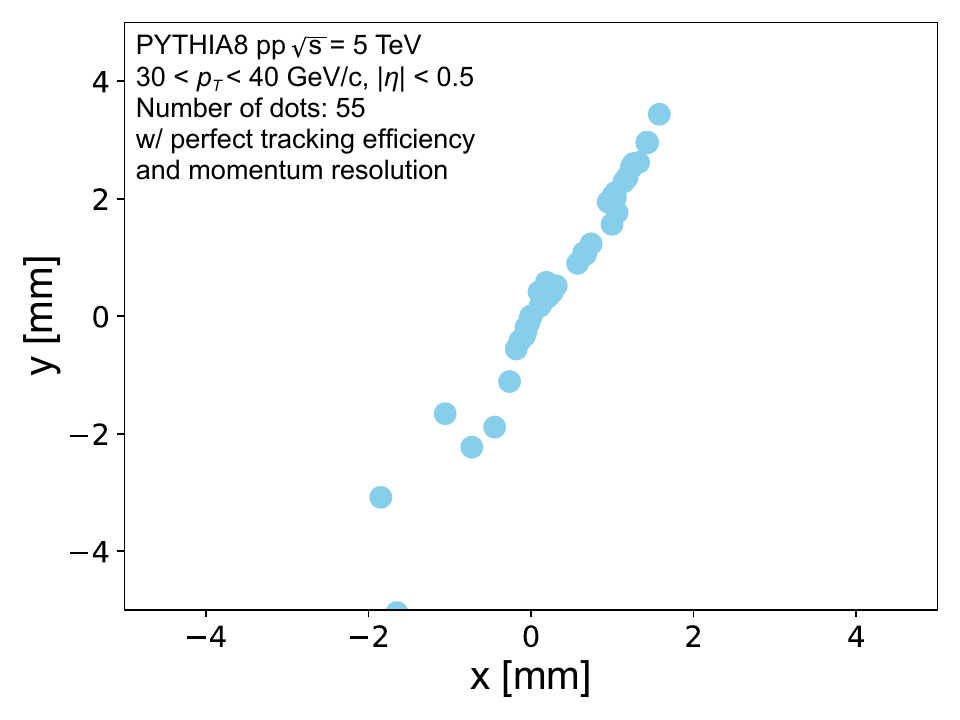}
\caption{Example 2D images of a $b$-jet with lines (left) and dots (right) with perfect tracking efficiency and momentum resolution.}
\label{fig:image_bjet}
\end{figure}

\begin{figure}[htb]
\includegraphics[width=0.49\textwidth]{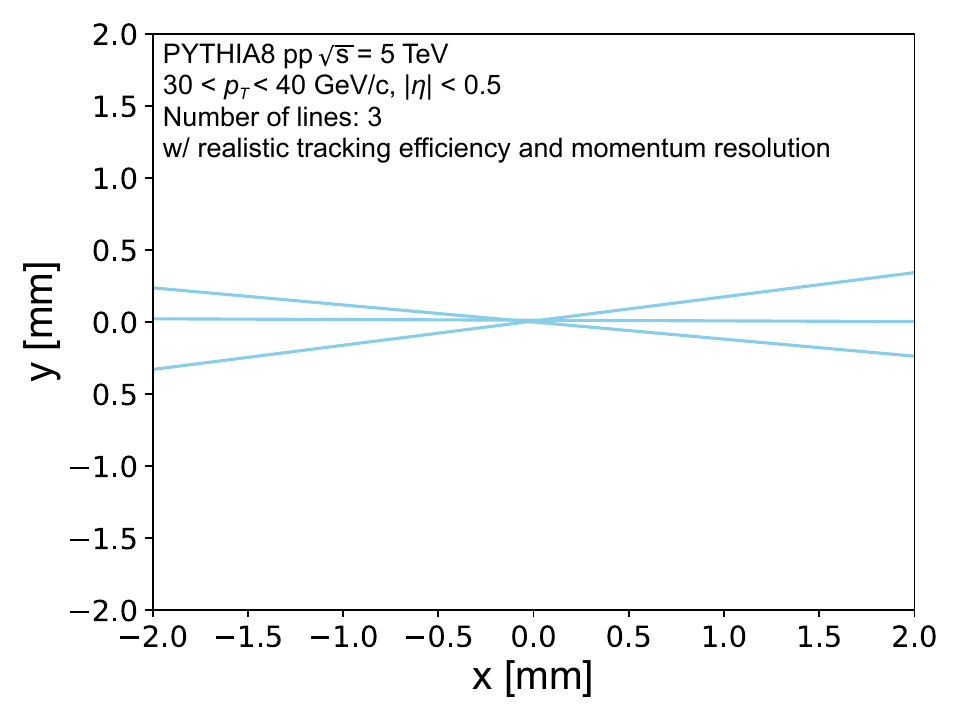}
\includegraphics[width=0.49\textwidth]{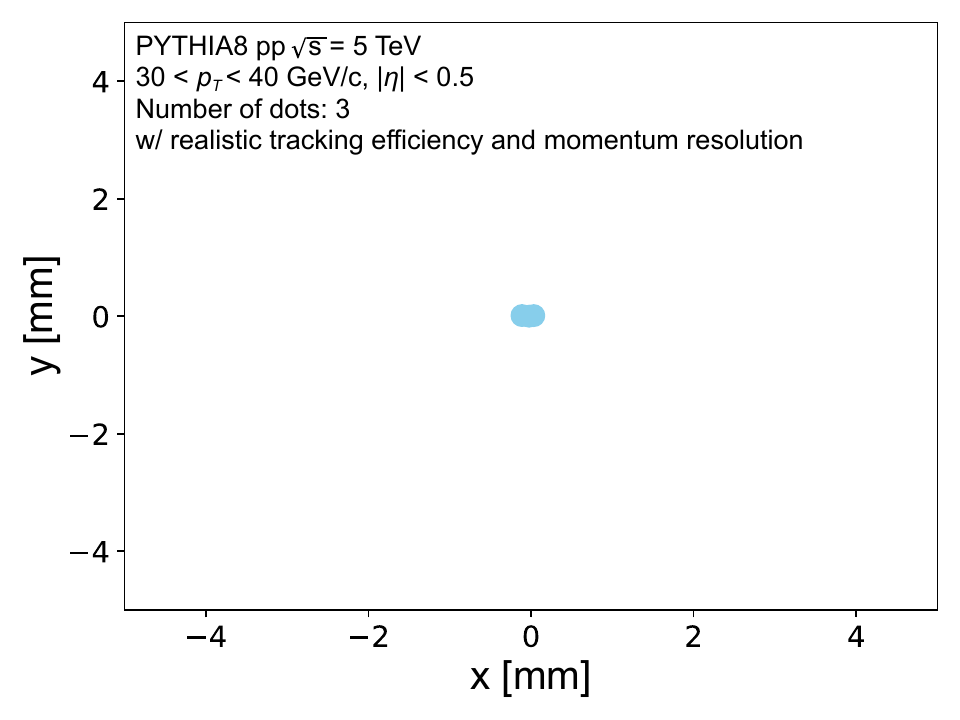}
\caption{Example 2D images of a light-jet with lines (left) and dots (right) with realistic tracking efficiency and momentum resolution.}
\label{fig:image_ljet_real}
\end{figure}

\begin{figure}[htb]
\includegraphics[width=0.49\textwidth]{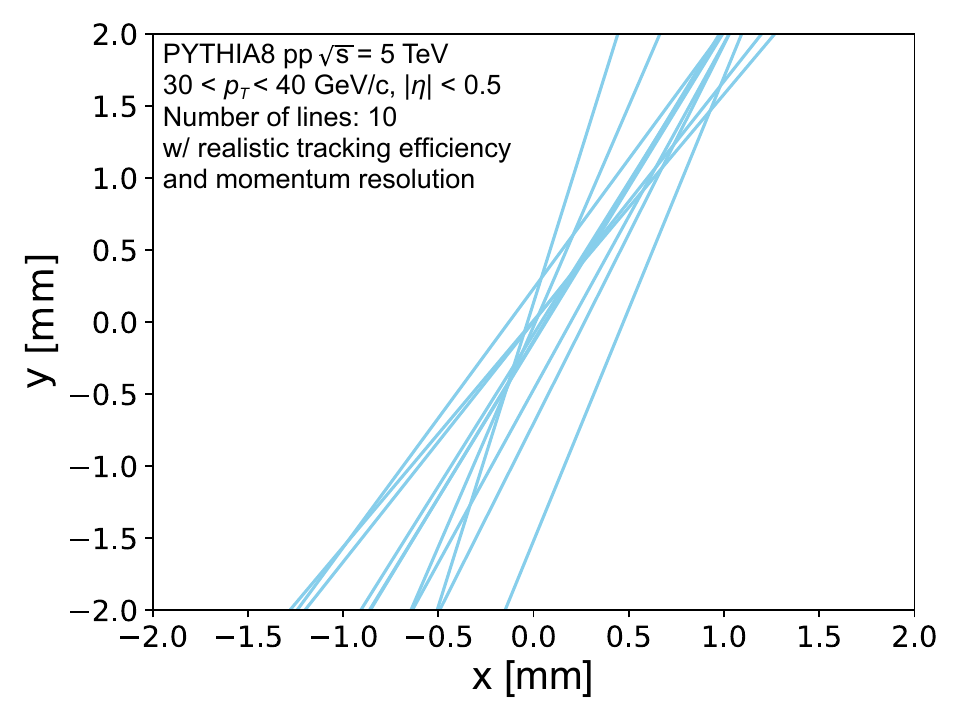}
\includegraphics[width=0.49\textwidth]{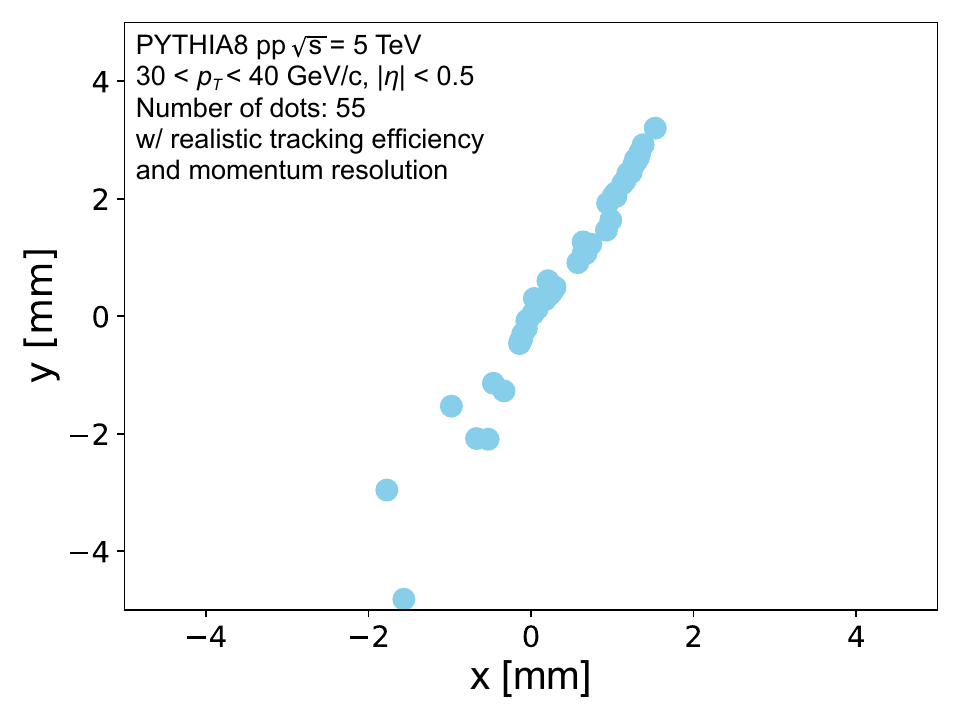}
\caption{Example 2D images of a $b$-jet with lines (left) and dots (right) with realistic tracking efficiency and momentum resolution.}
\label{fig:image_bjet_real}
\end{figure}

Figures~\ref{fig:image_ljet} and~\ref{fig:image_bjet} show example line (left) and dot (right) images of light-jet and $b$-jet with perfect tracking efficiency and momentum resolution.
Unlike the light-jet, several constituent charged particles are produced away from the primary vertex, so there are many dots of intersections.
Note that the radius of dots is set to 200~$\mu$m, which is similar to the typical resolution of a secondary vertex with a small number of tracks~\cite{ATLAS:2019wqx,ATLAS:2011cpj}.
In the case of light-jet, there are six crossing points due to particles produced from resonance decay, but the dots largely overlap due to the much shorter decay length (an order of fm) than the dot size.
In addition, there are more constituent charged particles in the $b$-jet than in the light-jet.
Figures~\ref{fig:image_ljet_real} and~\ref{fig:image_bjet_real} show example line (left) and dot (right) images of light-jet and $b$-jet with realistic tracking efficiency and momentum resolution, and there is little effect on the example 2D images.
We select a $2\times 2~\mathrm{mm}^{2}$ image size for line images to provide a closer look at particles produced from $b$-hadron decays, and a broader $5\times 5~\mathrm{mm}^{2}$ range is used for dot images to accommodate more intersections.
After producing the initial line and dot images, a preprocessing is applied.
Each image is resized to a fixed dimension of $128\times 128$ pixels, ensuring a uniform input size for the neural network.
This resizing facilitates consistent processing and reduces computational complexity.


\section{Neural Network Architecture}
\label{sec:nn}
\subsection{Model architecture}

\begin{figure}[htb]
\includegraphics[width=0.8\textwidth]{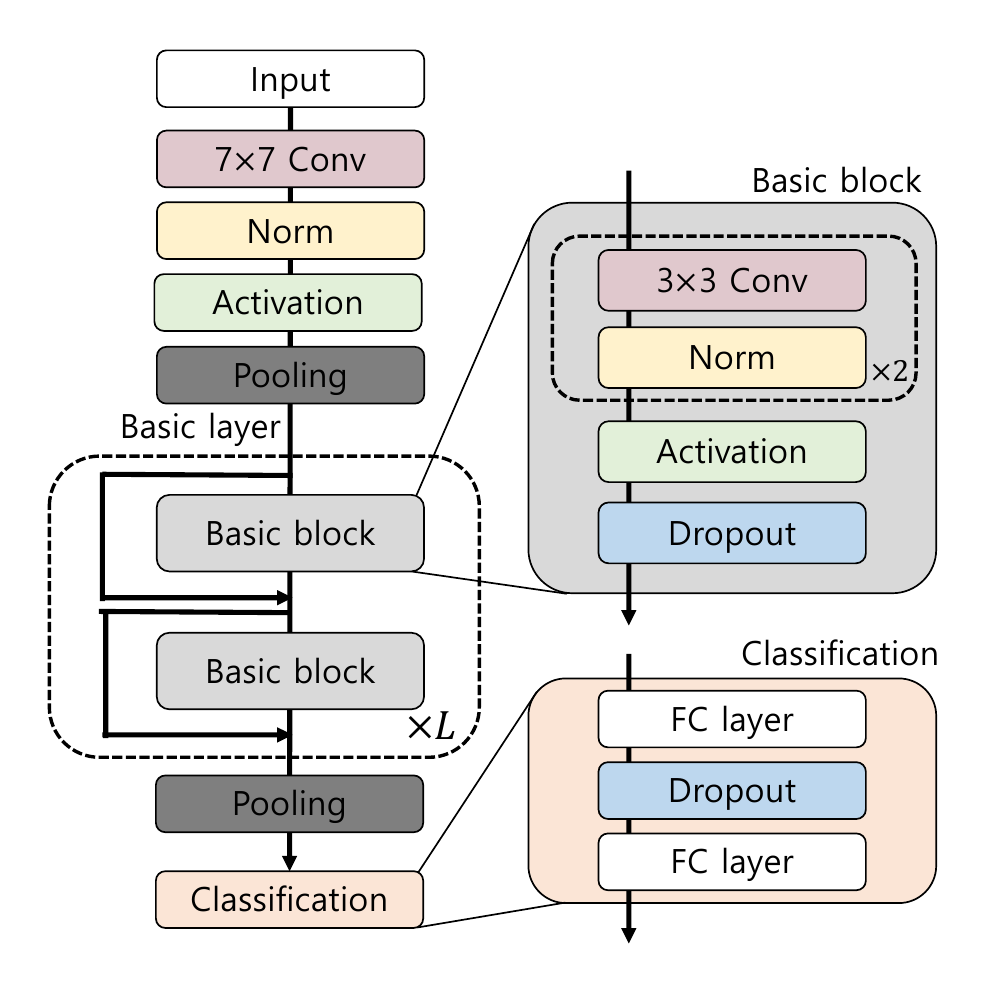}
\caption{The neural network architecture of ResNet-18 with dropout. }
\label{fig:nn_model}
\end{figure}

\begin{figure}[htb]
\includegraphics[width=0.4\textwidth]{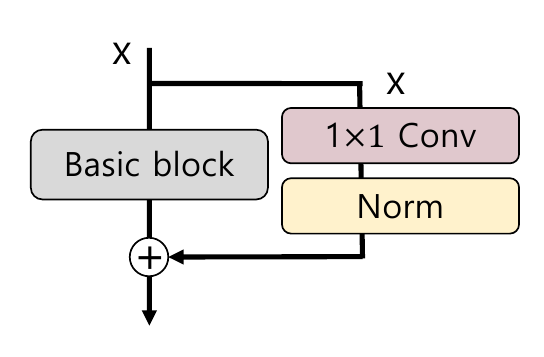}
\caption{The skip connection in the basic layer. This layer consists of a convolution layer with a filter size of $1 \times 1$ and batch normalization.}
\label{fig:nn_skip}
\end{figure}

The neural network model for the jet flavor tagging is shown in Fig.~\ref{fig:nn_model}, and arrows (or flow lines) indicate the sequential progression and direction of steps, guiding the process and elucidating dependencies between stages.  We select a residual neural network (ResNet-18)~\cite{He:2015wrn} with a dropout layer added to the convolutional neural network and fully connected layers to reduce overfitting. 
The dot and line images for jets undergo a convolution layer with filter sizes of $7 \times 7$, reducing their size by half and increasing the number of channels from 3 to 64.
Batch normalization, the RELU activation function~\cite{relu}, and max pooling are applied in the first block. 
The channels of the input image that pass through the convolutional neural network vary from 64 to 512.

The basic block consists of a convolution layer with a filter size of $3 \times 3$, followed by batch normalization, a RELU activation function, and dropout with a rate of 0.3.
The convolution and normalization repeat 2 times. 
Each basic block uses a skip connection, as shown in Fig.~\ref{fig:nn_skip}. 
The skip connection structure consists of a convolution layer with a filter size of $1 \times 1$ and batch normalization. 
The convolution layer is configured with a stride of 1 to maintain the original spatial dimensions. 
After passing through the basic block, the input is combined with the input that has passed through the skip connection. 
This technique helps preserve the gradient flow and enables deeper networks by mitigating the vanishing gradient problem commonly encountered in very deep networks.
In the basic layer, the basic block is used twice, and a residual connection is employed between the blocks. 
Additionally, this layer was repeated $L$ times, where $L$ was set to 4, the default value used in this study.
After the final basic layer, the input is passed through the adaptive average pooling layer. 
The classification consists of a multi-layer perceptron (MLP) with a fully connected layer and dropout with a rate of 0.4. 
Classification layers are often more prone to overfitting than convolutional layers due to their larger number of parameters. 
To mitigate overfitting, the dropout rate is increased to 0.4 in the fully connected layers, compared to the 0.3 dropout rate used in the convolutional layers.
The final output of the MLP has probabilities for three classes of jet flavor. 

\subsection{Training}

\begin{table}[h]
\centering
\caption{A summary of the hyperparameters used for training.}
\label{tab:parameter}
\begin{tabular}{cc}
\hline
\ Number of images per flavor & 20000 \\
\hline
\ Number of epochs & 50 \\
\ Early stopping & 20 epochs \\
\hline
Batch size & 1024 \\
\hline
\ Loss function & softmax cross-entropy \\
\hline
\ Optimizer & Adam~\cite{adam} \\
\hline
\ Learning rate scheduler & MultiStepLR \\
\ Learning rate & 0.001 -- 0.00001 \\
\hline
\end{tabular}
\end{table}

Table~\ref{tab:parameter} summarizes the hyperparameters used for the training.
We use 20000 jets for each flavor and \pt bin to prepare dot and line images, and the \pt bin follows $[20, 30, 40, 50, 60, 80, 100]$ in $\mathrm{GeV}/c$ unit.
The training and validation data are divided into a 6:4 ratio for usage.
In addition to the training and validation datasets, an independent test set consisting of 20000 images per flavor (60000 images in total) was prepared, entirely separate from the training and validation samples. All performance results reported in this paper are based on this test set, ensuring that the evaluation is unbiased by the training or validation data.
The Adam optimizer~\cite{adam} is used for model optimization with a batch size of 1024 over 50 epochs, and the loss for each epoch is calculated using softmax cross-entropy loss. 
%
The learning rate scheduler ``MultiStepLR'' is used because users can optimize the learning rate in epochs. The learning rate starts at 0.001 and decreases by a factor of 0.1 in the range of epochs from 5 to 15. It decreases again by a factor of 0.1 after epoch 15, reaching a minimum value of 0.00001. The training procedure is stopped when the valid loss does not decrease for 20 epochs.


\section{Results}
\label{sec:results}
The ResNet-18 with dropout tagger predicts the probability that a jet belongs to the $b$-jet, $c$-jet, and light-jet classes. To use the model for $b$-jet tagging, these probabilities are combined into a single score $D_{b}$ by following the recent study using deep learning algorithms from the ATLAS Collaboration~\cite{ATLAS:2022rkn}.
It is defined as 
\begin{equation}
   \\\\ D_b = \log\frac{p_b}{(1 - f_c)p_l + f_cp_c},
\end{equation}
where $p_b$, $p_c$, and $p_l$ are jet flavor probabilities, and $f_c$, as a hyper-parameter, determines the weights of $c$-jet and light-jet, thereby influencing $c$-jet and light-jet rejection. 
The parameter is set to $f_c=0.017$ for the tagger using dot images, while an $f_c$ value of 0.06 is used for the tagger using line images.
These $f_c$ parameter values are determined through an optimization process that maximizes the rejection of $c$-jets and light-jets at a fixed 85\% working point in terms of $b$-jet efficiency.

\begin{figure}[htb]
\includegraphics[width=0.7\textwidth]{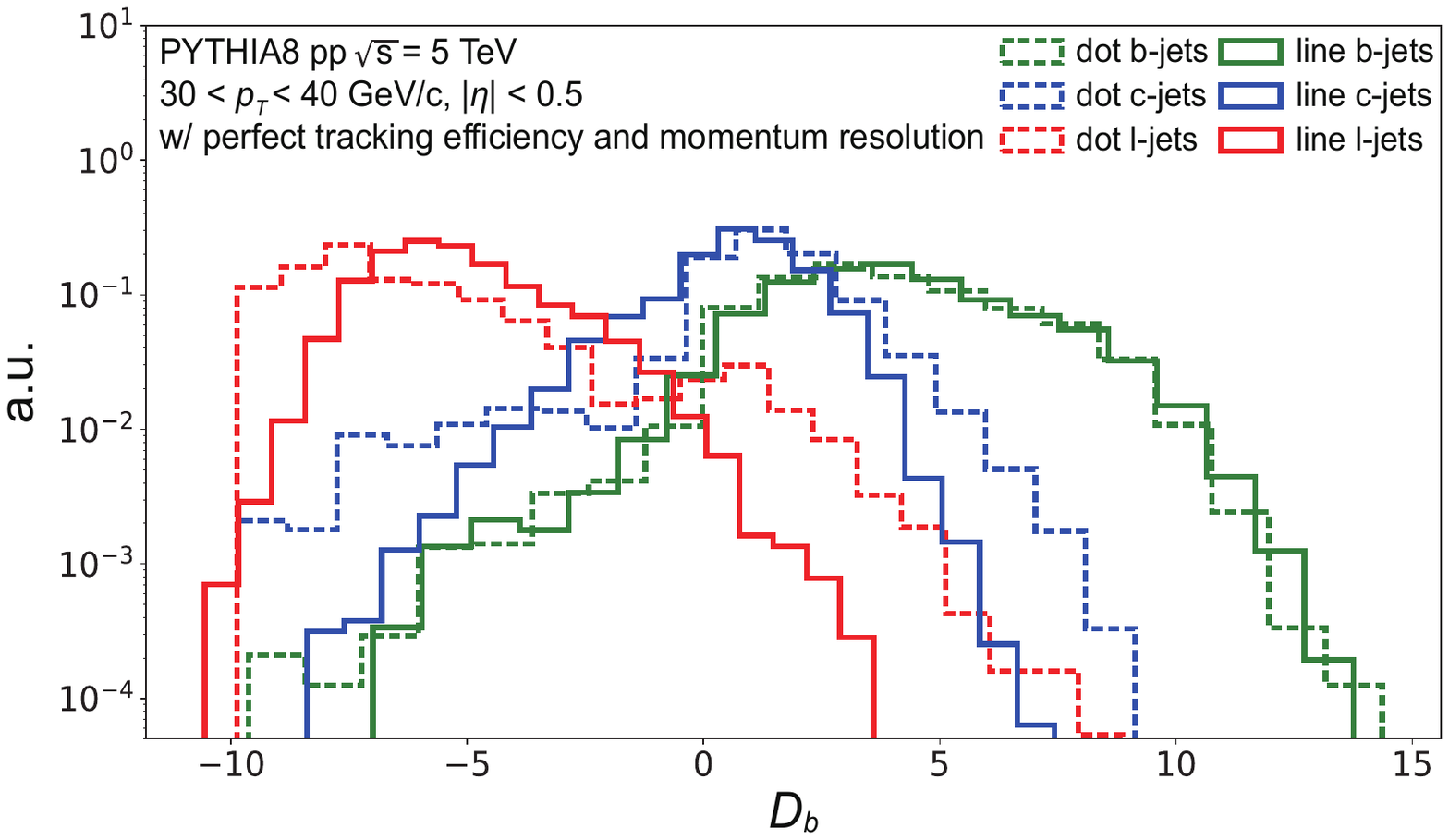}
\includegraphics[width=0.7\textwidth]{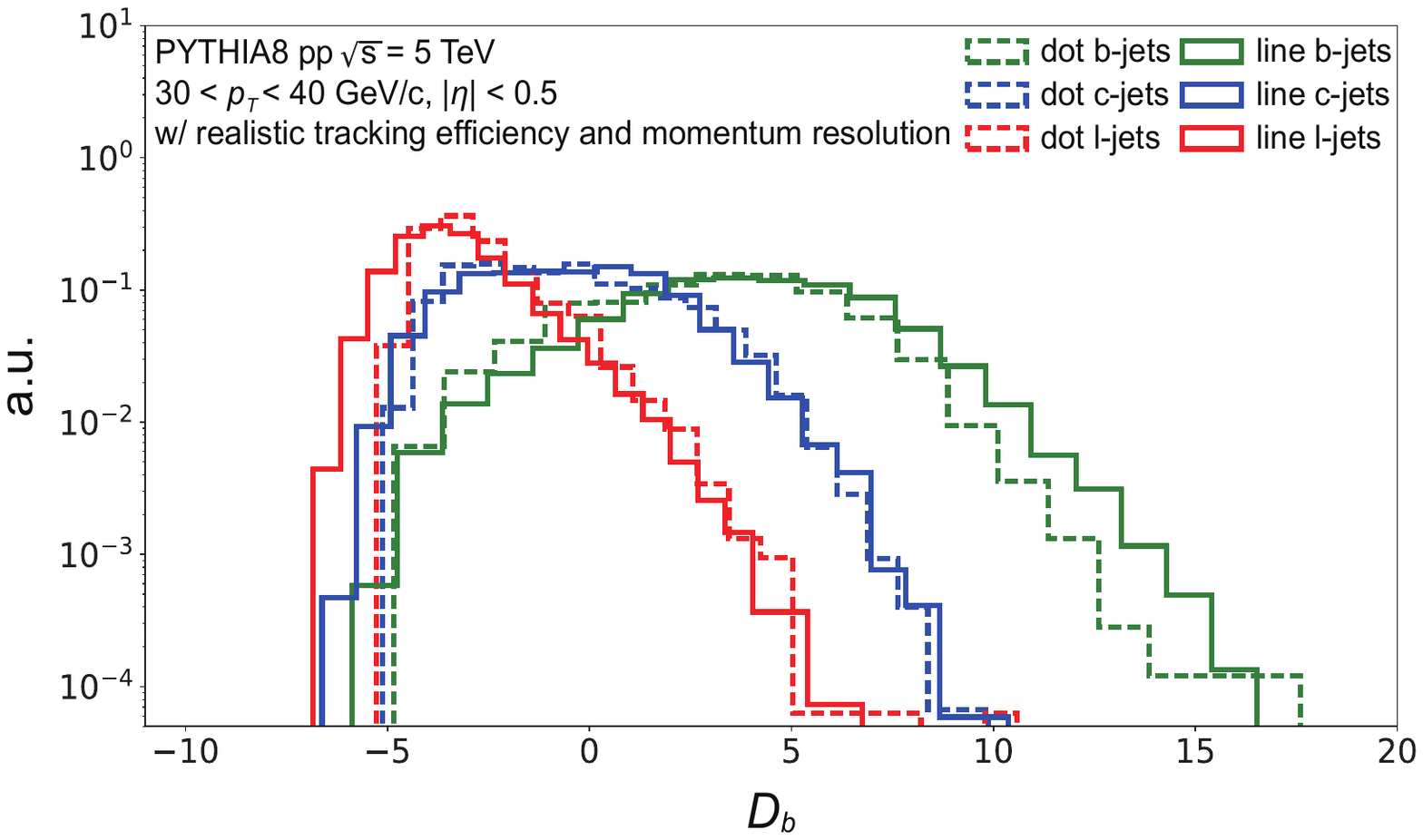}
\caption{Comparison of $b$-jet tagging discriminant ($D_b$) distributions between two taggers using dot and line images for jets in $30<p_{T}<40~\mathrm{GeV}/c$ and $|\eta|<0.5$. The upper (lower) panel represents the result with perfect (realistic) tracking efficiency and momentum resolution.}
\label{fig:d_b}
\end{figure}

Figure~\ref{fig:d_b} shows $b$-jet tagging discriminant ($D_b$) distributions for two taggers using dot and line images for jets in $30<p_{T}<40~\mathrm{GeV}/c$ and $|\eta|<0.5$, and the upper (lower) panel represents the result with perfect (realistic) tracking efficiency and momentum resolution.
In both cases of perfect and realistic conditions, the distributions from the two taggers, dots and lines, for $ b$-jets exhibit a similar shape. 
However, the distributions from the tagger using dot images for light-jet and $c$-jet are shifted towards higher values than those from the tagger using line images, resulting in better performance with the tagger using line images.
When comparing perfect and realistic conditions, the overall shape of $D_b$ changes, but a clear discrimination is still evident in the positive $D_b$ region.


\begin{figure}[htb]
\includegraphics[width=1\textwidth]{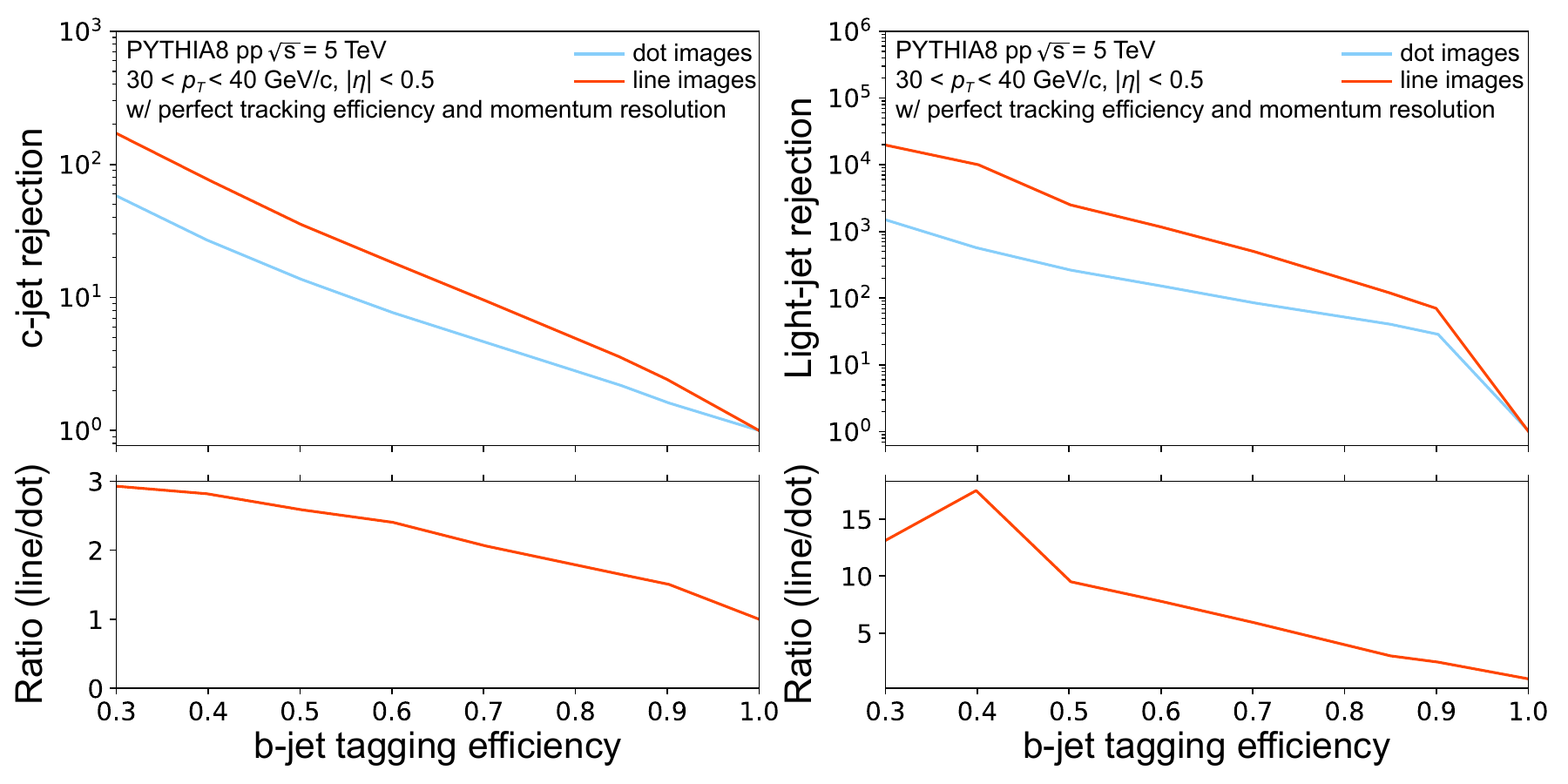}
\caption{The $c$-jet (left) and light-jet rejection as a function of $b$-jet tagging efficiency for jets in $30 < p_{T} < 40~\mathrm{GeV}/c$ and $|\eta|<0.5$ with perfect tracking efficiency and momentum resolution. The bottom plots show the ratio between two taggers.}
\label{fig:rejection}
\end{figure}

\begin{figure}[htb]
\includegraphics[width=1\textwidth]{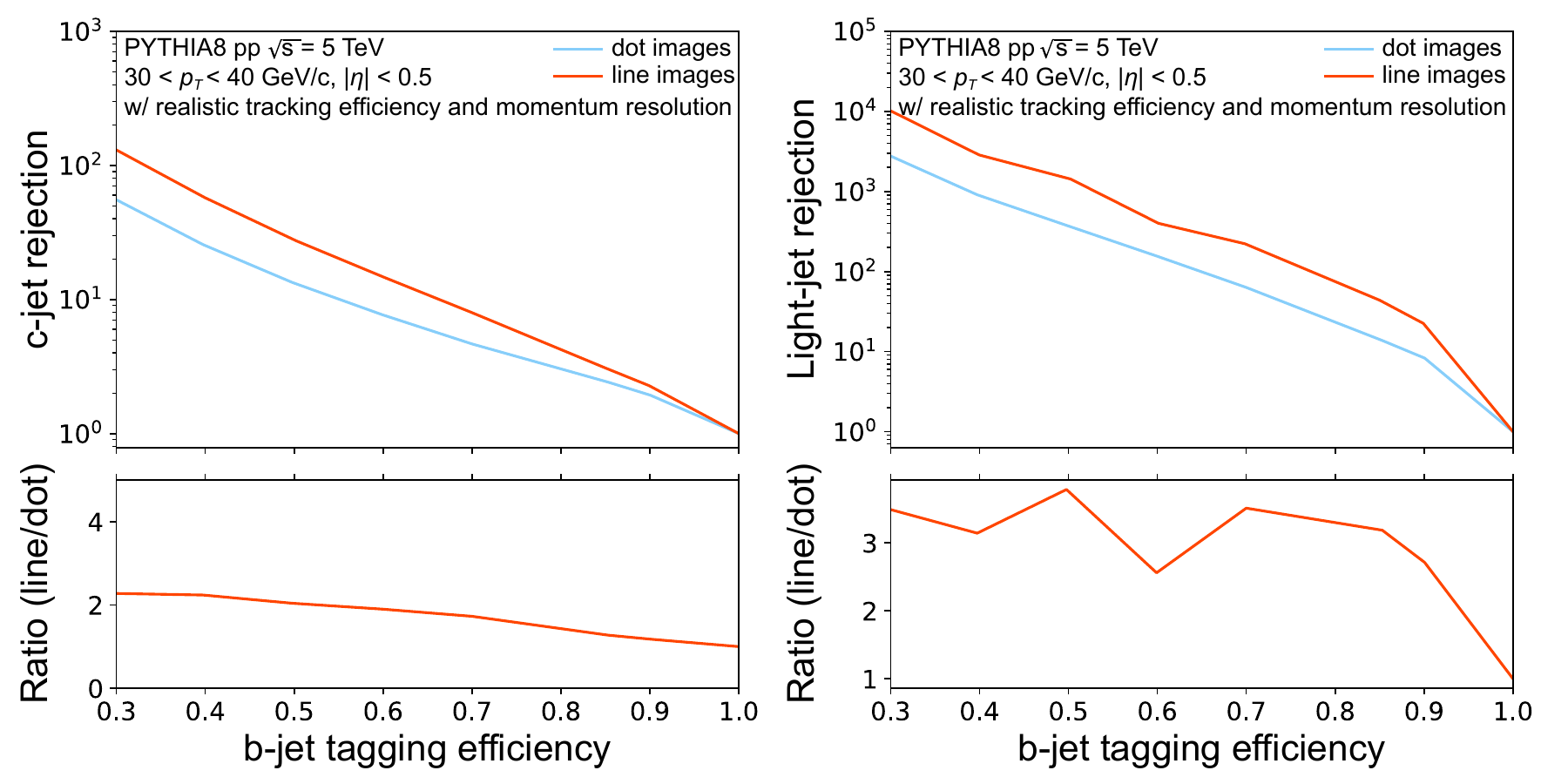}
\caption{The $c$-jet (left) and light-jet rejection as a function of $b$-jet tagging efficiency for jets in $30 < p_{T} < 40~\mathrm{GeV}/c$ and $|\eta|<0.5$ with realistic tracking efficiency and momentum resolution. The bottom plots show the ratio between two taggers.}
\label{fig:rejection_real}
\end{figure}

\begin{figure}[htb]
\includegraphics[width=1\textwidth]{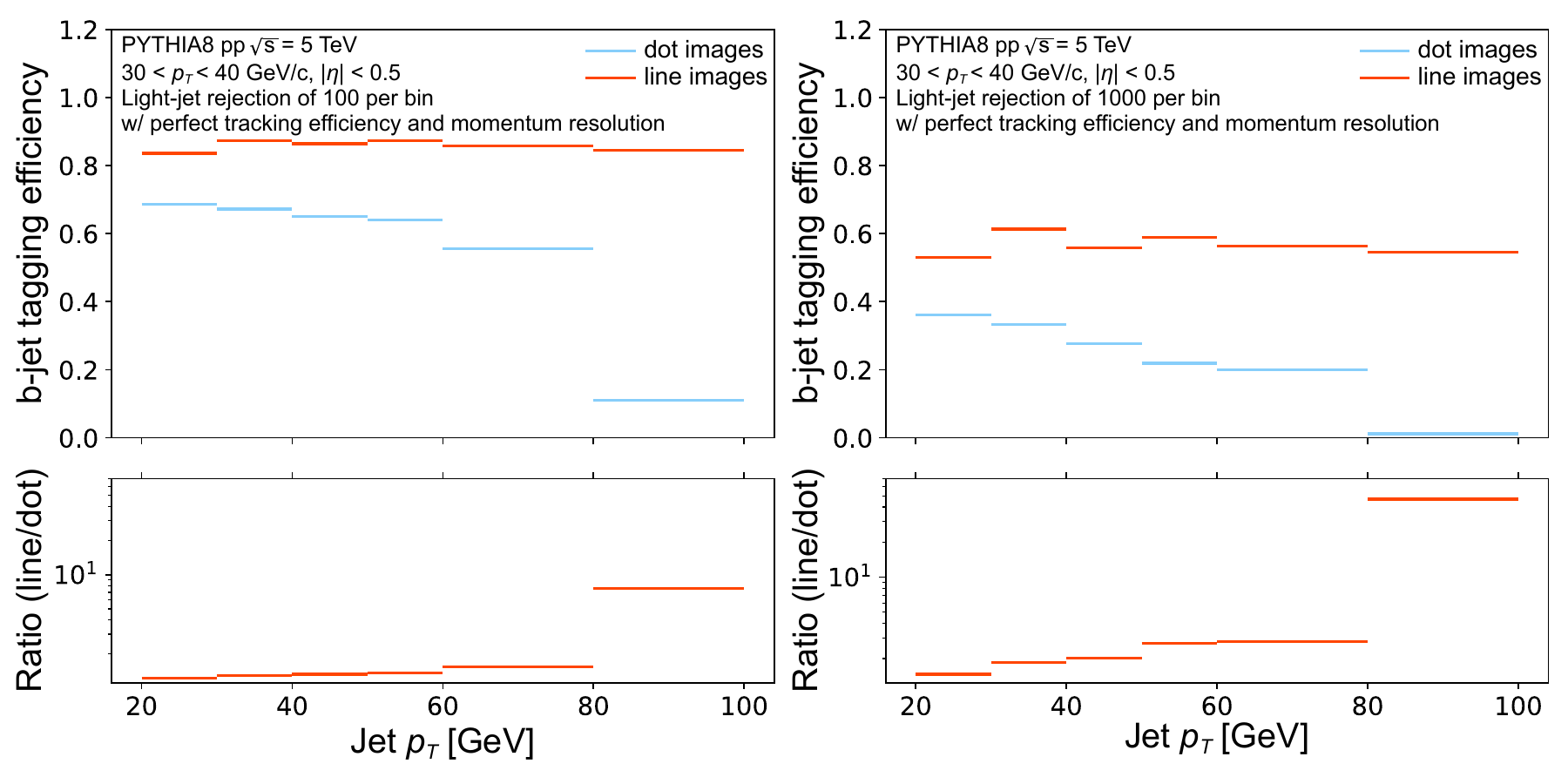}
\caption{The $b$-jet tagging efficiency as a function of jet $p_{T}$ from two taggers with a fixed light-jet rejection of 100 (left) or 1000 (right) in each bin with perfect tracking efficiency and momentum resolution.}
\label{fig:effi}
\end{figure}

\begin{figure}[htb]
\includegraphics[width=1\textwidth]{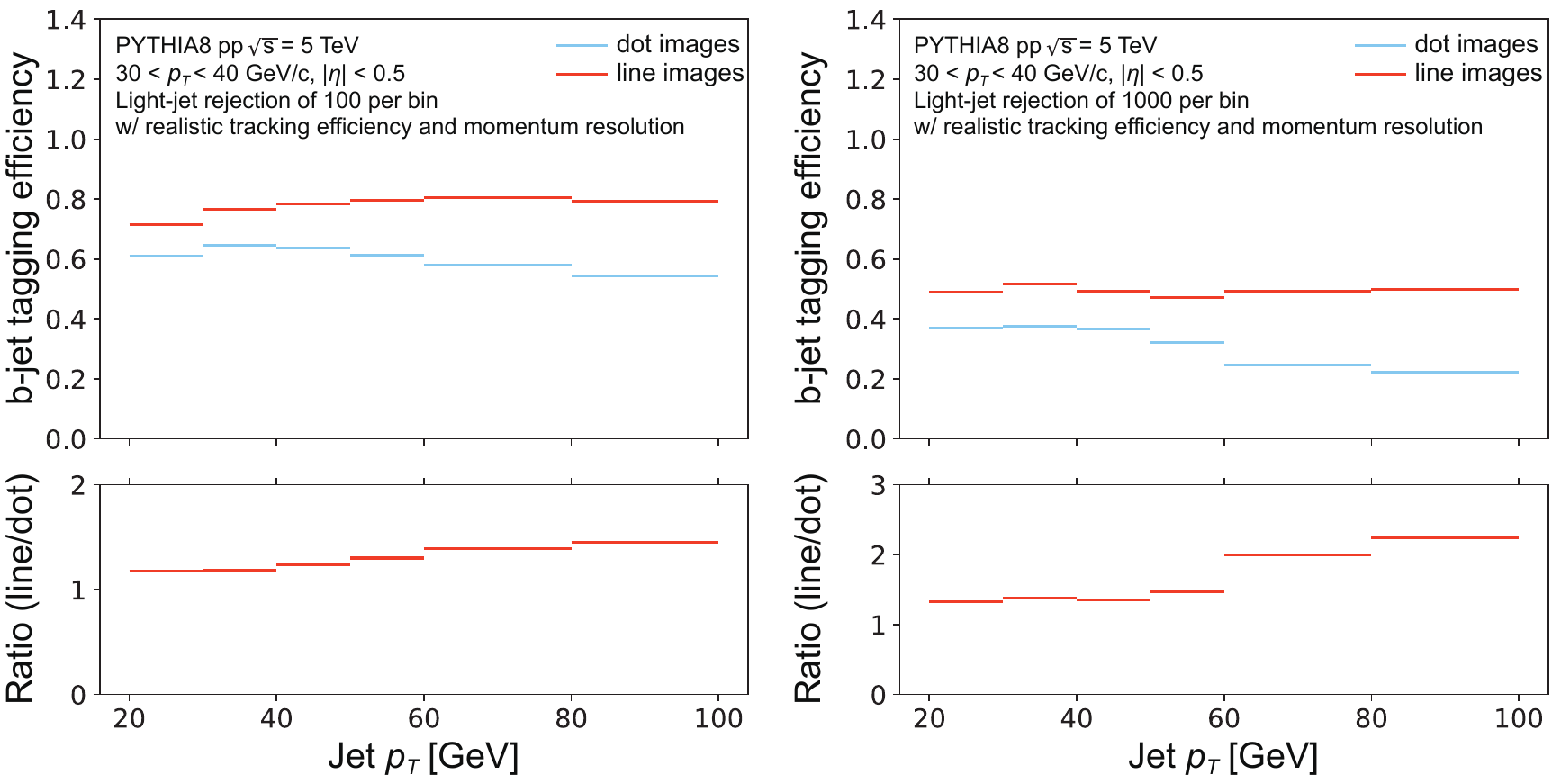}
\caption{The $b$-jet tagging efficiency as a function of jet $p_{T}$ from two taggers with a fixed light-jet rejection of 100 (left) or 1000 (right) in each bin with real tracking efficiency and momentum resolution.}
\label{fig:effi_real}
\end{figure}

The performance of $b$-tagging is evaluated based on its ability to reject $c$-jets and light-jets at different $b$-jet tagging efficiencies. 
The rejection rates of $c$-jet (left) and light-jet (right) for jets in $30<p_{T}<40~\mathrm{GeV}/c$ and $|\eta|<0.5$ are shown in Figs.~\ref{fig:rejection} and~\ref{fig:rejection_real}.
The ratio of performance from the two taggers is presented in the bottom panels.
Under perfect detector conditions, the tagger using line images exhibits significantly improved performance in rejecting $ c$-jets and light-jets, while maintaining the same $b$-jet tagging efficiency. 
Specifically, at a $b$-jet tagging efficiency of around 40\%, the line image shows an improvement of approximately 3 times for $c$-jet rejection and about 17 times for light-jet rejection compared to the dot image. 
The performance difference gradually decreases as the $b$-jet tagging efficiency increases.
Under realistic detector conditions, a slight improvement is observed in the performance with dot images, whereas the performance with line images shows a modest deterioration. Consequently, the discrepancy between the two taggers is reduced in this context.

The performance of $b$-jet tagging algorithms may vary with jet \pt.
As jet \pt increases, the number of fragmentation particles also increases significantly and is more collimated, leading to a more crowded environment within the jet.
This increase in particle multiplicity complicates the task of distinguishing individual particle tracks, which is essential for identifying and categorizing particles originating from different processes, including heavy-flavor decays.
The $b$-jet tagging efficiency with two taggers is evaluated across various \pt ranges.
As shown in Table~\ref{tab:parameter}, we employ a separate training procedure for different \pt bins using 20000 images for each tagger, flavor, and \pt bin.

To evaluate the tagger performance with varying jet \pt, the $b$-jet tagging efficiencies for a fixed light jet rejection of 100 (left) or 1000 (right) per bin across different jet \pt values are shown in Figs.~\ref{fig:effi} and~\ref{fig:effi_real}.
In both cases of the detector condition, the tagger using line images shows a relatively stable $b$-jet tagging efficiency of 80--90\% at the entire jet \pt range for a 100 light-jet rejection, which is a better performance than the tagger using dot images. 
Especially in the high \pt region of $80 < p_{T} < 100~\mathrm{GeV}/c$, the most significant difference is observed, with the tagger using dot images showing a very low efficiency of approximately 10\%.
At high \pt, the increased particle multiplicity leads to more intersections even in light jets, making jet classification only with intersections more difficult.
However, this improves under realistic detector conditions with a 3--5\% momentum resolution, since the finite resolution makes the crossing point appear broader. As a result, it becomes less sensitive to spurious crossing points not originating from b-decays, effectively leading to a single merged crossing point.
With the selection of 1000 light-jet rejections, the $b$-jet tagging efficiency from the tagger using line images is about 60\% and remains almost independent of \pt.
Similar to the performance results of light and $c$-jet rejection, the performance difference between the two taggers becomes smaller when considering realistic tracking efficiency and momentum resolution.
In our analysis, the models were trained separately in narrow \pt bins [20, 30, 40, 50, 60, 80, and 100 GeV/$c$], which largely mitigates biases arising from within-bin \pt shape discrepancies. However, since the jet \pt spectrum in actual data may differ significantly from the PYTHIA simulation used here, the relative performance between line-image and dot-image methods observed in this study could vary when applied to real data. Future validation using reweighted \pt distributions or data-driven \pt control samples is therefore necessary to confirm robustness in experimental settings.

\begin{figure}[htb]
\includegraphics[width=1\textwidth]{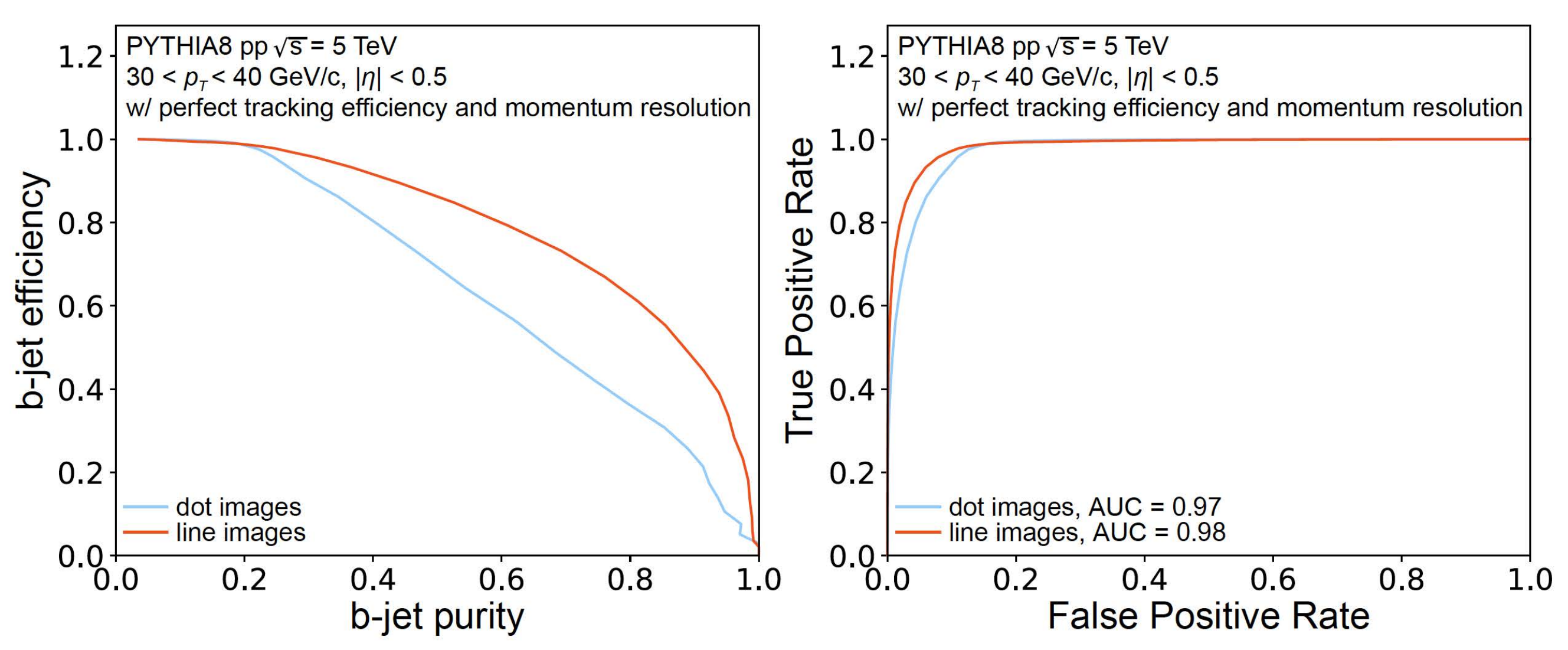}
\includegraphics[width=1\textwidth]{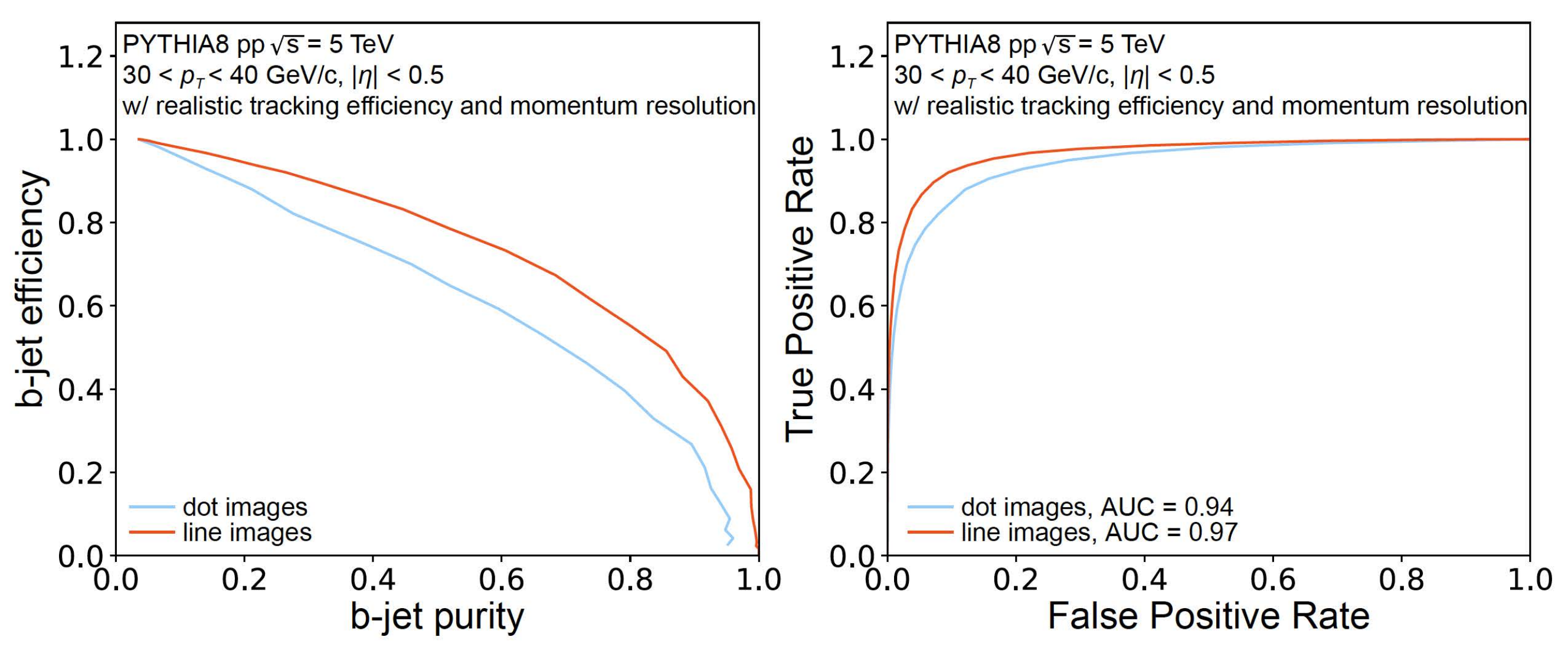}
\caption{The purity and efficiency (left) and ROC curves (right) for the $b$-jet classification.  The upper (lower) panel represents the result with perfect (realistic) tracking efficiency and momentum resolution.}
\label{fig:classification}
\end{figure}

Additionally, we evaluate the $b$-jet tagging performance in terms of purity and the receiver operating characteristic (ROC) curve, considering the relative contributions of $b$-, $c$-, and light-jets, with ratios of $1:2:25$ based on the PYTHIA8 settings.
Figure~\ref{fig:classification} shows the performance curves for jets in $30<p_T<40~\mathrm{GeV}/c$ and $|\eta|<0.5$, and the upper (lower) panel represents the result with perfect (realistic) tracking efficiency and momentum resolution.
Under ideal detector conditions, at a $b$-jet efficiency of 60\%, the purity of the dot image tagger is approximately 60\%, while the purity of the line image tagger is around 80\%.
The purity at the same 60\% efficiency slightly decreases under realistic detector conditions.
In addition, a selection for high purity ($>90\%$) with reasonable efficiency ($\sim50\%$) can be achieved using the line image tagger.
Again, this higher purity is due to the line image’s ability to more effectively capture and represent the spatial relationships and intersection patterns of tracks within the jet, leading to more accurate classification.
The area under the curve (AUC) of the receiver operating characteristic (ROC) curve is computed for each origin class using a one-versus-all classification approach.
In the ROC curve, the AUC for the line image is 0.98 (0.97), compared to 0.97 (0.94) for the dot image under ideal (realistic) detector conditions.
Both taggers show good AUC values, but the slight difference makes a larger performance difference in the efficiency vs. purity curve due to the small yield of $b$-jet than $c$- and light-jet.
Note that the true positive rate of both taggers with the ideal detector condition converges to 1 at a 0.2 false positive rate, and it changes to 0.7 false positive rate with the realistic detector condition.

\begin{figure}[htb]
\includegraphics[width=0.49\textwidth]{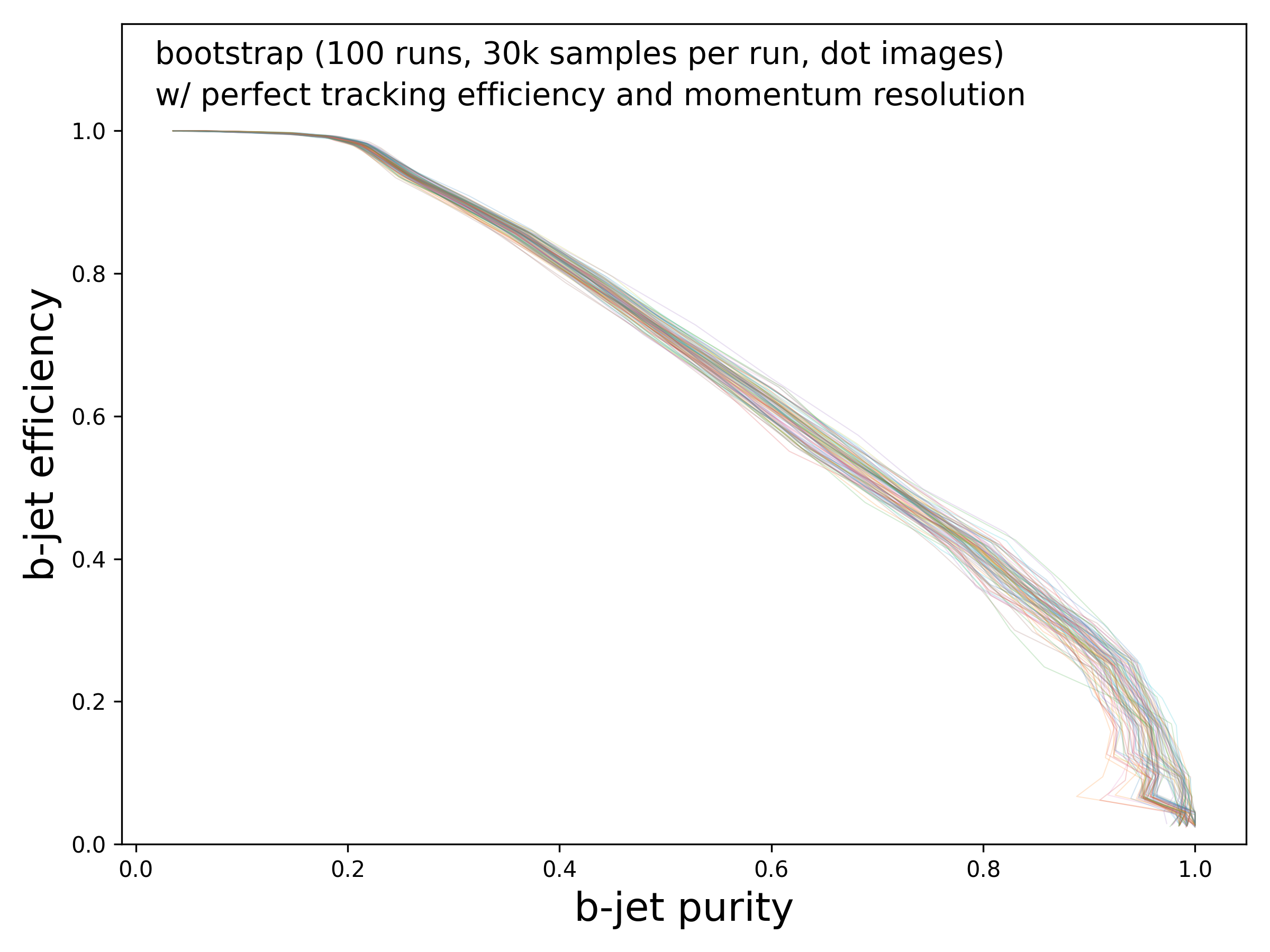}
\includegraphics[width=0.49\textwidth]{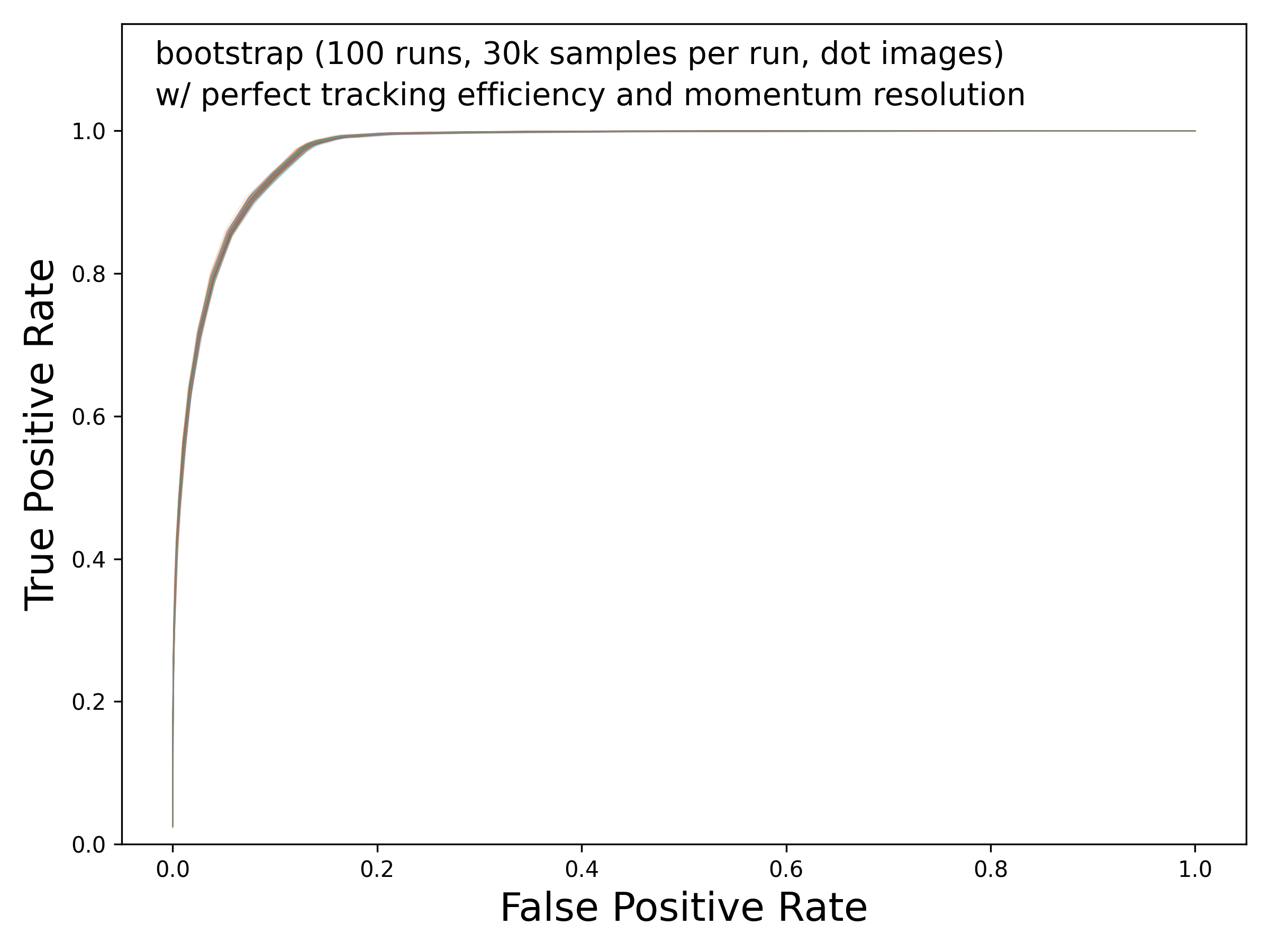}
\includegraphics[width=0.49\textwidth]{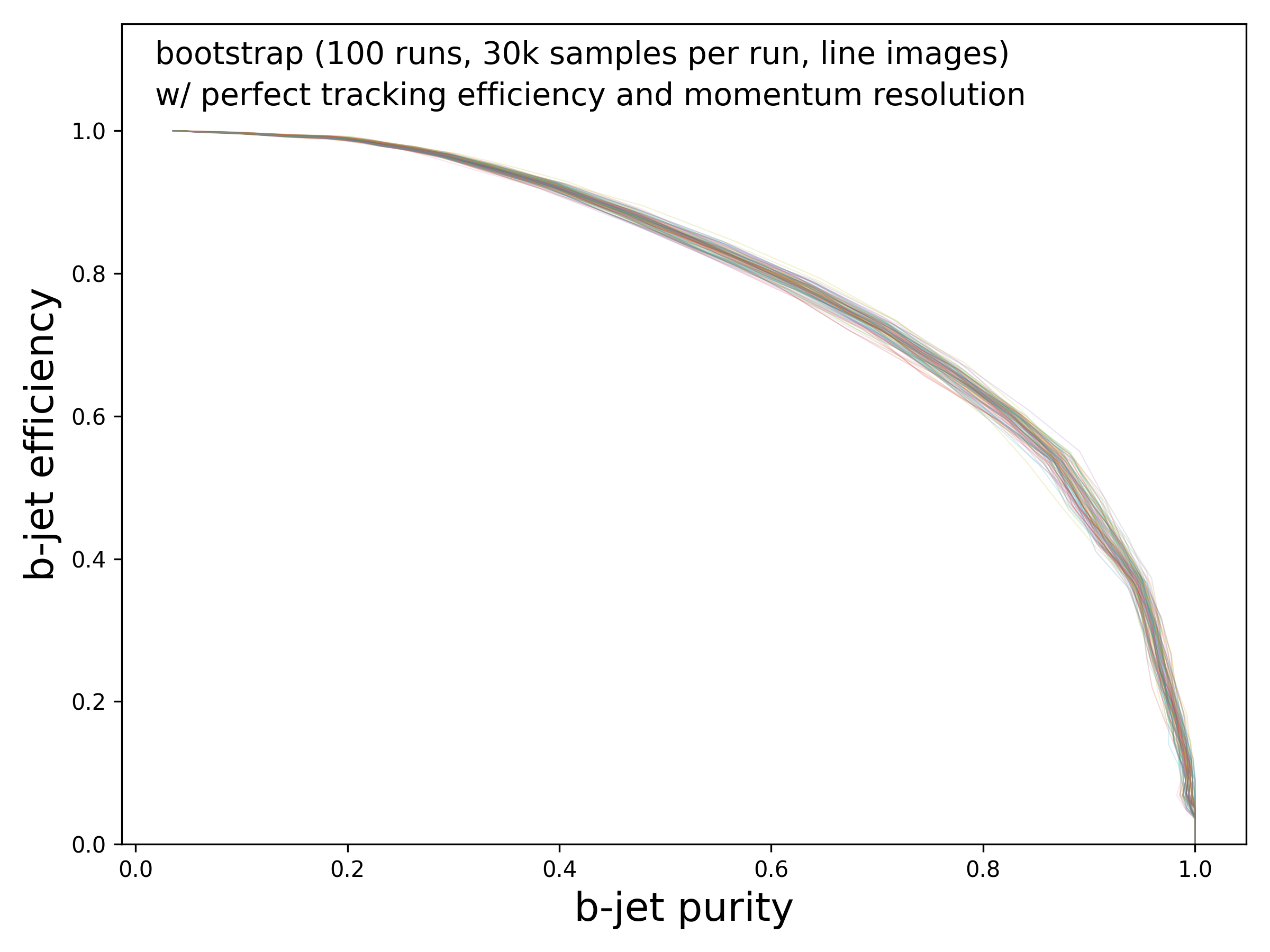}
\includegraphics[width=0.49\textwidth]{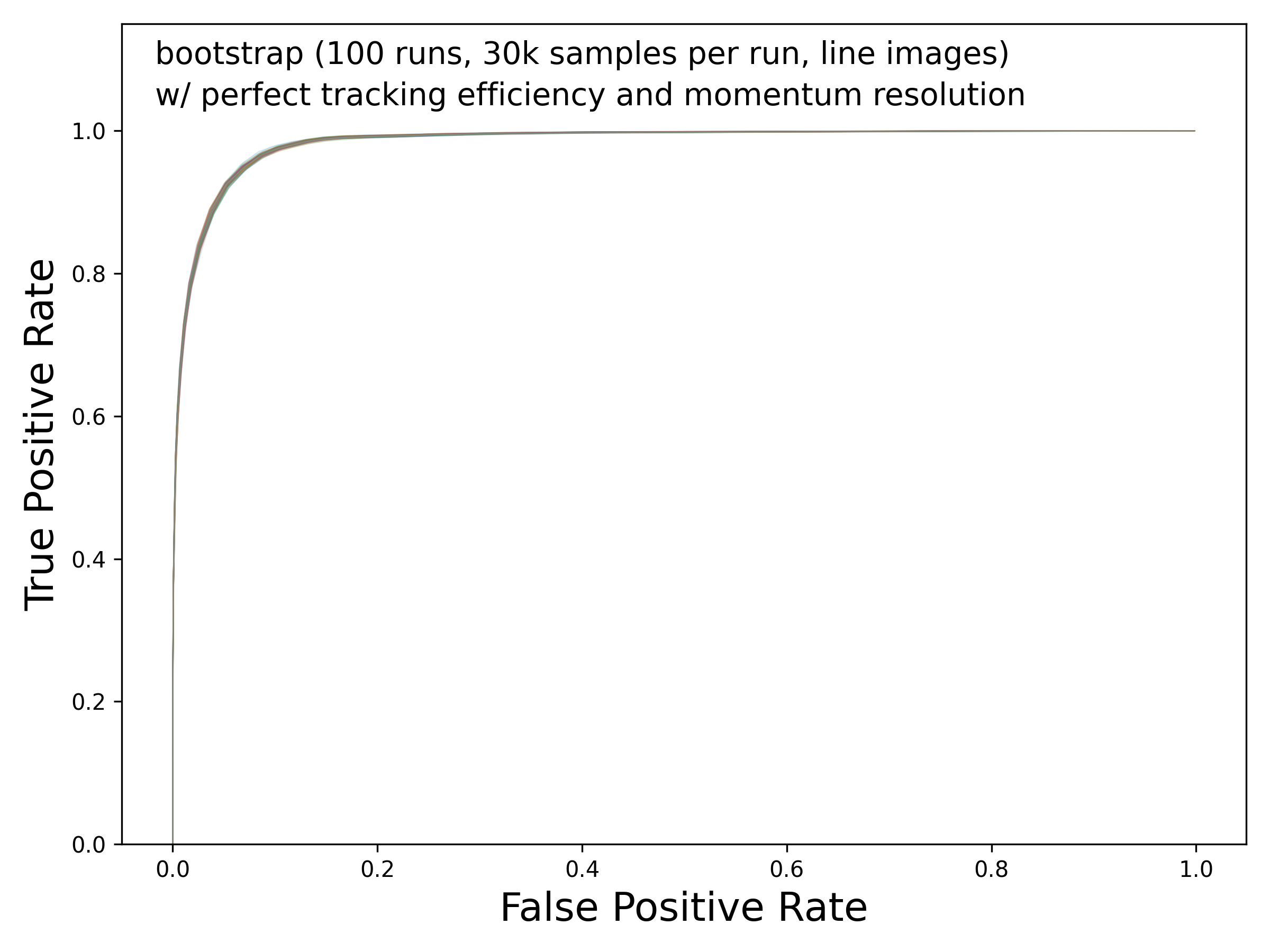}
\caption{Purity and efficiency (left) and ROC curves (right) for $b$-jet classification, obtained from 100 bootstrap test samples, each containing 10,000 images per jet flavor. The top panels show results using dot images, while the bottom panels correspond to line images, both under ideal detector conditions.}
\label{fig:bootstrap}
\end{figure}

To evaluate the uncertainty in the $b$-jet tagging performance, we employed a bootstrap method. From a total of 60000 test sample images (20000 per flavor), we randomly selected 30000 images (10000 per flavor) to compute the $b$-jet tagging performance as shown in Fig.~\ref{fig:classification}. This procedure was repeated 100 times to assess the variation in performance. The results are summarized in Fig.~\ref{fig:bootstrap}: the top panels correspond to the use of dot images, while the bottom panels represent line images. All results are based on the perfect detector condition, and similar trends were observed under realistic detector conditions as well.

For dot images, the variation in efficiency at a fixed purity was generally below 5\% in the purity vs. efficiency plots, with slightly larger variations observed in the high-purity, low-efficiency region. This increased variation appears to stem from whether the randomly selected light-jet images contain weak decay products, which significantly affects performance at high purity and low efficiency. In contrast, the results based on line images showed more stable performance with smaller variations.

The ROC curves on the right-hand side exhibited variations of less than 1\% across the 100 bootstrap samples. The comparatively larger variation seen in the purity vs. efficiency plots can be attributed to the flavor imbalance in the sample, where the relative $l$-jet:$c$-jet:$b$-jet ratio was set to 25:2:1. As a result, variations in $l$-jet rejection are amplified in the final performance due to the overwhelming abundance of $l$-jets compared to $b$-jets.

\section{Summary}
\label{sec:summary}
In this paper, we explore jet flavor tagging by utilizing image recognition from the field of computers.
The images are produced with the track information within the jet, assuming both an ideal tracking system and realistic conditions of tracking efficiency and momentum resolution, based on the expected performance of the ALICE ITS3. 
They include line images drawn along the directions of the particles and dot images marking the intersections of these lines.
We utilized the ResNet-18 model and added dropout. 
We find that the tagger using line images shows a better performance than the tagger using dot images, particularly with the line image tagger demonstrating approximately 2 times and 6 times better $c$-jet and light-jet rejection, respectively, compared to the dot image tagger for jets within $30 < p_{T} < 40~\mathrm{GeV}/c$ at a $b$-jet tagging efficiency of 70\% under the ideal detector condition, and the performance difference between two tagger reduces considering realistic detector conditions.
Additionally, at high \pt, within the range $80 < p_{T} < 100~\mathrm{GeV}/c$, the line image tagger maintains a $b$-jet tagging efficiency of 80\%, while the efficiency of the dot image tagger rapidly drops to around 10\%, making classification impossible for the ideal detector condition.
But it is improved with a finite momentum resolution, resulting in a broader size of crossing points.
Additionally, we examine the purity and ROC curves based on the relative contribution of different jet flavors in PYTHIA8.
The line image tagger shows about 70\% of $b$-jet tagging efficiency and purity, which is optimistic for further development of this method for actual analysis of experimental data.


\begin{acknowledgments}

H.G. Jang and S.H. Lim acknowledge support from the National Research Foundation of Korea (NRF) grant funded by the Korean government (MSIT) under Contract No. RS-2025-00554431 and NRF-2008-00458. We also acknowledge technical support from KIAF administrators at KISTI.  

\end{acknowledgments}

\bibliographystyle{unsrt}   
\bibliography{reference}

\pagebreak
\appendix
\section{Additional figures}
\counterwithin{figure}{section}
\setcounter{figure}{0}

\begin{figure}[htb]
\includegraphics[width=0.7\textwidth]{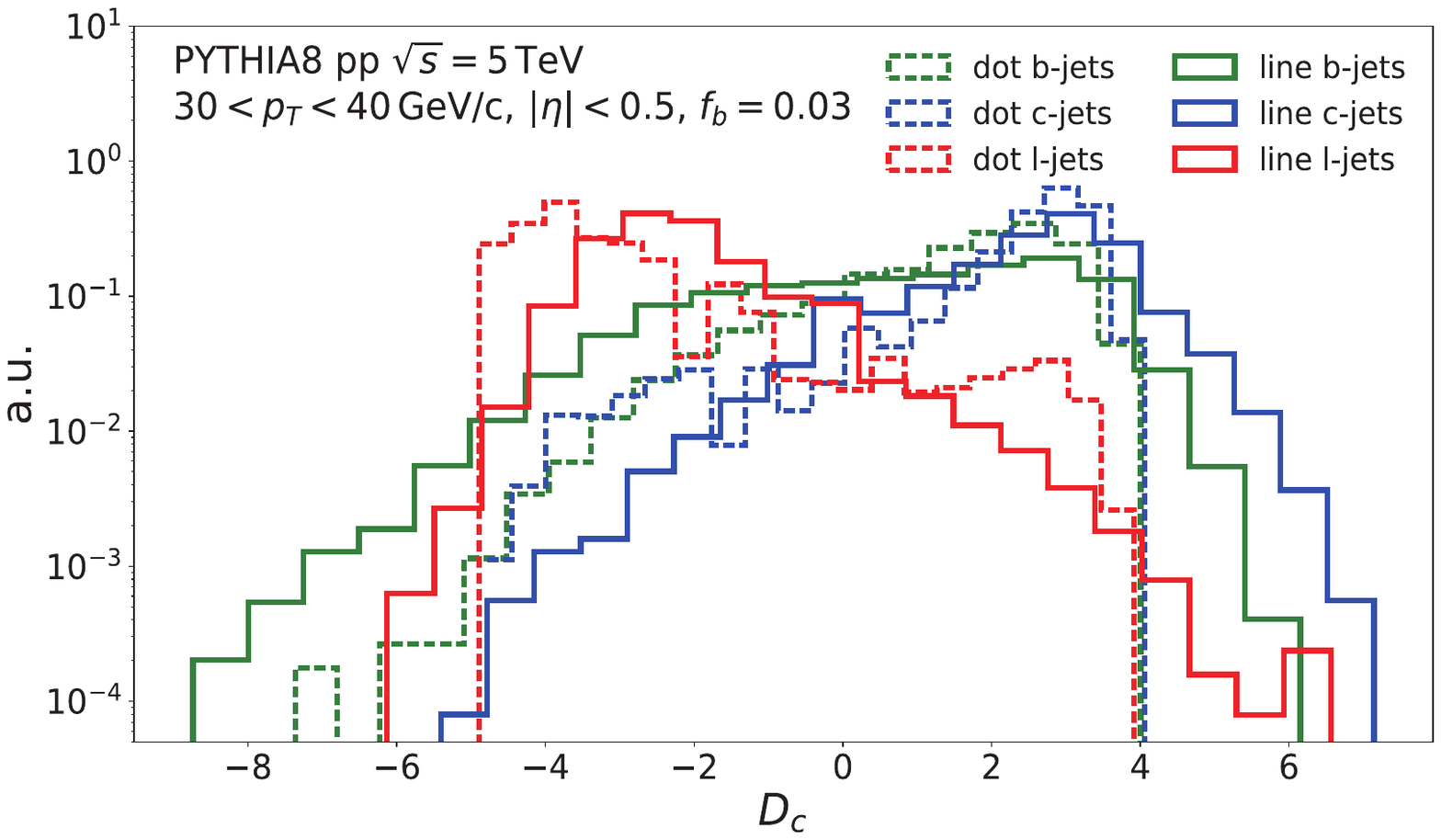}
\caption{Comparison of $c$-jet tagging discriminant ($D_c$) distributions between two taggers using dot and line images for jets in $30<p_{T}<40~\mathrm{GeV}/c$ and $|\eta|<0.5$.}
\label{fig:d_c}
\end{figure}

\begin{figure}[htb]
\includegraphics[width=1\textwidth]{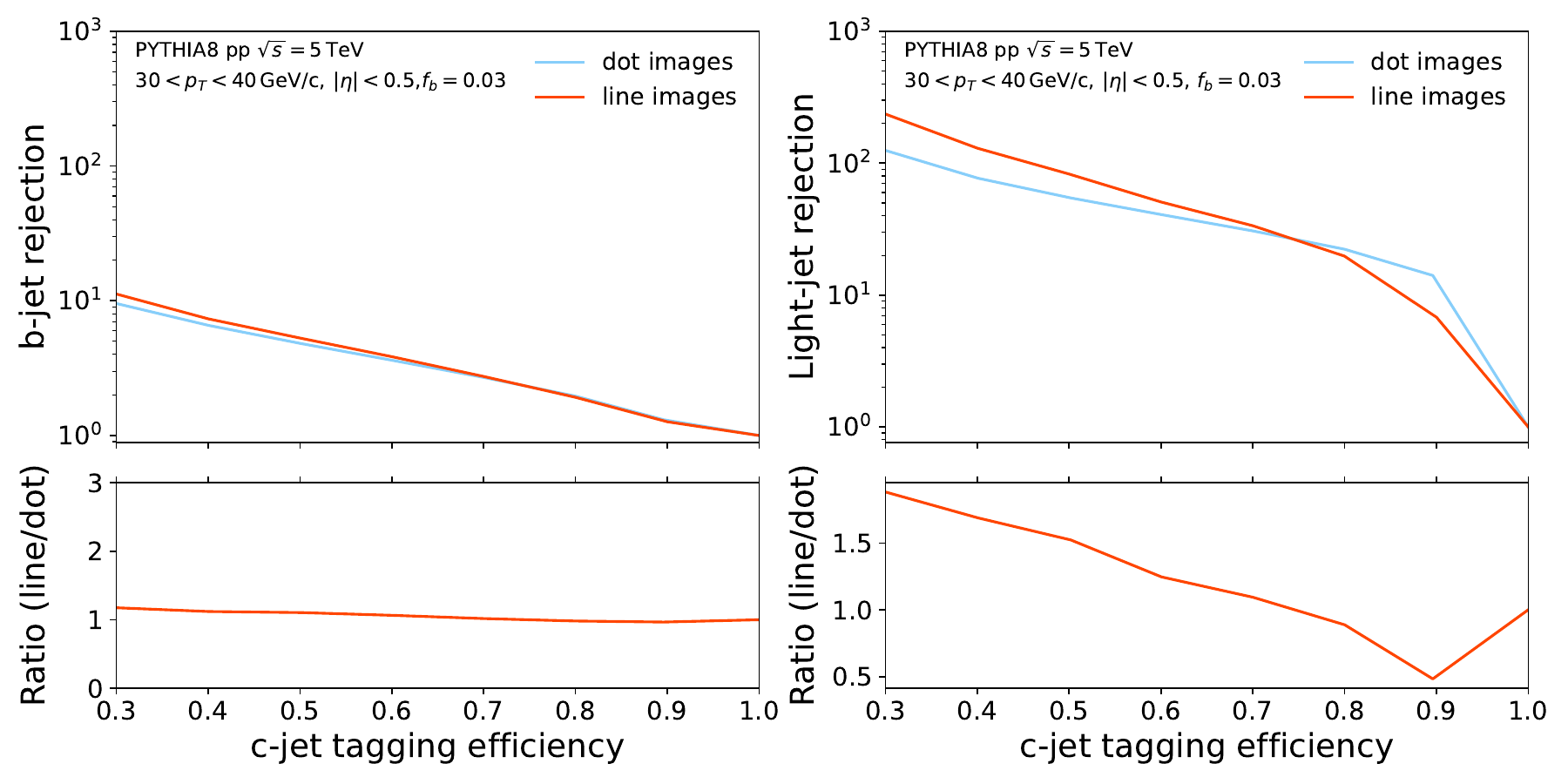}
\caption{$b$-jet (left) and light-jet rejection as a function of $c$-jet tagging efficiency for jets in $30 < p_{T} < 40~\mathrm{GeV}/c$ and $|\eta|<0.5$. The bottom plots show the ratio between two taggers.}
\label{fig:cjet_rejection}
\end{figure}

\begin{figure}[htb]
\includegraphics[width=1\textwidth]{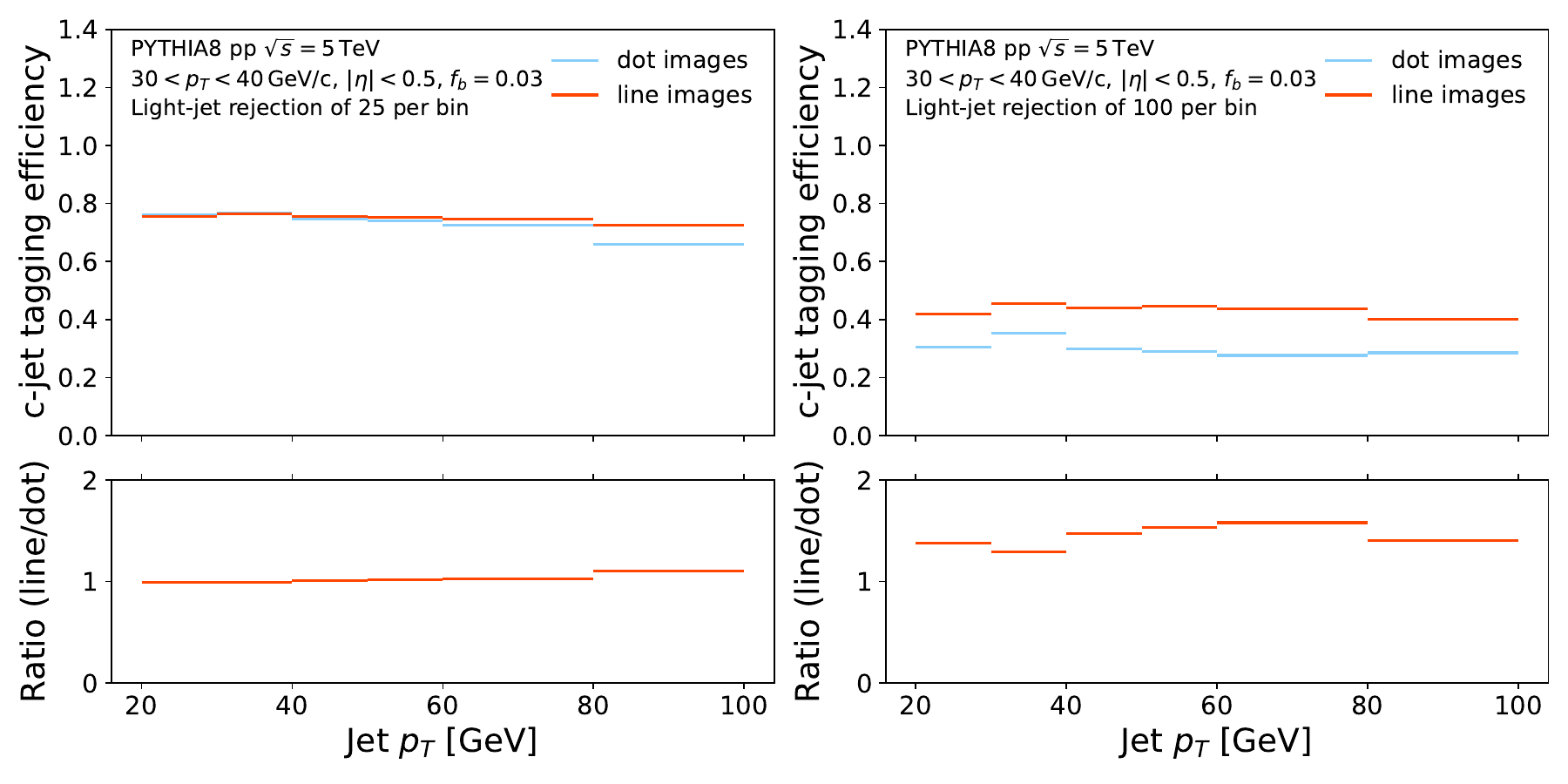}
\caption{The $c$-jet tagging efficiency as a function of jet $p_{T}$ from two taggers with a fixed light-jet rejection of 25 (left) or 100 (right) in each bin.}
\label{fig:cjet_effi}
\end{figure}

\begin{figure}[htb]
\includegraphics[width=1\textwidth]{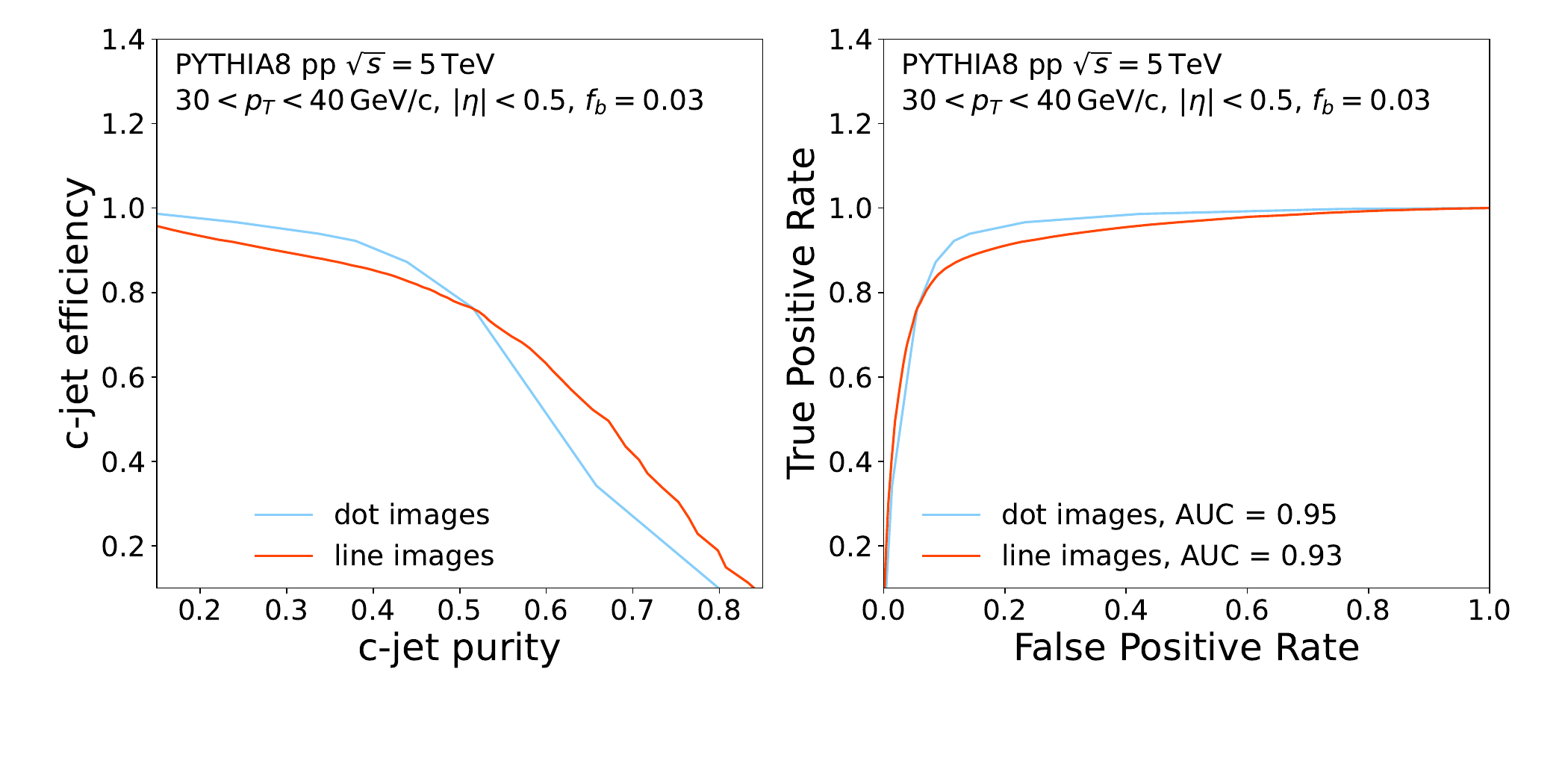}
\caption{The purity and efficiency (left) and ROC curves (right) for the $c$-jet classification.}
\label{fig:cjet_classification}
\end{figure}

Here, we present additional figures (Figs.~\ref{fig:d_c}--\ref{fig:cjet_classification}) for the $c$-jet tagging performance, specifically for the case of tracking efficiency and perfect momentum resolution. It is based on the combined score for $c$-jet defined as 
\begin{equation}
   \\\\ D_c = \log\frac{p_c}{(1 - f_b)p_l + f_b p_b},
\end{equation}
and the weights of $b$-jet and light-jet $f_b=0.03$ is selected to optimize $c$-jet and light-jet rejection. 

\end{document}